\documentclass[twocolumn,showpacs,preprintnumbers,amsmath,amssymb]{revtex4}
\usepackage{graphicx}% Include figure files
\usepackage{dcolumn}% Align table columns on decimal point
\usepackage{bm}% bold math

\newcommand{\bra}[1]{\langle #1|}
\newcommand{\ket}[1]{|#1\rangle}
\newcommand{\braket}[2]{\langle #1|#2\rangle}

\newtheorem{corollary}{Corollary}
\newtheorem{theorem}{Theorem}
\newtheorem{definition}{Definition}

\begin{document}

%\preprint{APS/123-QED}

\title{A Robust Semidefinite Programming Approach to the Separability Problem}% Force line breaks with \\

\author{Fernando G. S. L. Brand\~ao}
\email{fgslb@ufmg.br}
\author{Reinaldo O. Vianna}
\email{reinaldo@fisica.ufmg.br}
\affiliation{Universidade Federal de Minas Gerais - Departamento  de F\'{\i}sica\\
Caixa Postal 702 - Belo Horizonte - MG -  Brazil - 30.123-970}
\date{\today}

%\date{}% It is always \today, today,
             %  but any date may be explicitly specified

\begin{abstract}
We express the optimization of entanglement witnesses for arbitrary bipartite states in terms of a class of convex optimization problems known as Robust Semidefinite Programs (RSDP). We propose, using well known properties of RSDP, several new sufficient tests for separability of mixed states. Our results are then generalized to multipartite density operators.

\end{abstract}

\pacs{03.67.Mn}% PACS, the Physics and Astronomy
                             % Classification Scheme.
%\keywords{Suggested keywords}%Use showkeys class option if keyword
                              %display desired
\maketitle

\section{INTRODUCTION}

\hspace{0.2 cm} Entanglement, first noticed by Einstein, Podolsky, and Rosen \cite{1}, is at the heart of quantum mechanics. Quantum teleportation, superdense coding and cryptography \cite{2} are achieved only when one deals with inseparable states. Thus, the determination and quantification of entanglement in a composite quantum state is one of the most important tasks of quantum information theory. In the past years a great deal of effort have been made in order to obtain the characterization of separable bipartite mixed states \cite{3}. A finite-dimensional bipartite density operator $\rho_{AB} \in {\cal B}(H_{A} \otimes H_{B})$ (the Hilbert space of bounded operators acting on $H_{A}\otimes H_{B}$) is separable iff it can be written as a convex sum of separable pure states: 
\begin{equation}
\rho_{AB} = \sum_{i=1}p_{i} \ket{\psi_{i}}_{AA}\bra{\psi_{i}}\otimes \ket{\phi_{i}}_{BB}\bra{\phi_{i}}
\end{equation}
where ${\cal f}p_{i}{\cal g}$ is a probability distribution and $\ket{\psi_{i}}_{A}$, $\ket{\phi_{i}}_{B}$ are vectors belonging to Hilbert spaces $H_{A}$ and $H_{B}$, respectively. Despite the simplicity of this definition, none operational necessary and sufficient criterion have been found for the separability problem until now. Moreover, it was showed by Gurvits \cite{4} that this problem is NP-HARD. Therefore, we should not expect to find a polynomial-time algorithm which determines for any state $\rho_{AB}$, with certainty, if it is possible to decompose it in the form of equation $(1)$.

A particularly useful concept is that of entanglement witness (EW). According to \cite{5}, an operator $\rho_{AB}$ is entangled iff there exists a self-adjoint operator $W \in {\cal B}(H_{A} \otimes H_{B})$ which detects its entanglement, i.e., such that $Tr(W \rho_{AB}) < 0$ and $Tr(W\sigma_{AB}) \geq 0$ for all $\sigma_{AB}$ separable. This condition follows from the fact that the set of separable states is convex and closed in ${\cal B}(H_{A} \otimes H_{B})$. Therefore, as a conclusion of the Hahn-Banach theorem, for all entangled states there is a linear functional which separates it from this set. Unfortunately, it is not known how to construct EW in a canonical way and in polynomial time for every entangled state. Actually, since such method would solve the separability problem, it can not exist as long as the strong conjecture $P \neq NP$ is true. 

In this paper, we show that the search of EW for arbitrary mixed states is indeed NP-HARD. We introduce, in the context of quantum information, a class of convex optimization problems known as Robust Semidefinite Programs (RSDP), whose NP-hardness in most of the cases was already proved. This family generalizes the important Semidefinite Programs (SDP), which have been increasingly used in quantum information problems (see,  for example,   \cite{9} and \cite{10}). 

The paper is organized as follows. In section 2 we briefly recall the definition of EW and define our concept of optimal entanglement witness; in section 3 we state the basic facts about Robust Semidefinite Programs, express the optimization of EW as a RSDP and provide a first approximation in terms of SDP for the problem, which yields  a new sufficient criterion of separability; in section 4 we parametrize all possible approximations of the RSDP in terms of a multiplier matrix, reducing the search space of approximation scenarios; in section 5 we generalize our results to multipartite states. Finally, in section 6, we present our conclusions and suggest directions for further research.

\section{OPTIMAL ENTANGLEMENT WITNESS}

A hermitian operator $W \in {\cal B}(H_{A}\otimes H_{B})$ is an entanglement witness if \cite{5} \cite{6}:
\begin{enumerate}
\item $_{A}\bra{\psi}\otimes _{B}\bra{\phi}W\ket{\phi}_{B}\otimes \ket{
\psi}_{A}\geq 0$, for all states $\ket{\psi} \in H_{A}$ and $\ket{\phi} \in H_{B}$.
\item $W$ has at least one negative eigenvalue.
\item $Tr(W) = 1$.
\end{enumerate}
Condition one assures that $Tr(W\sigma_{AB}) > 0$ for all separable states $\sigma_{AB}$. Condition two implies that $Tr(WP) < 0$ at least for one entangled state, for example,  the projector on the eigenspace associated with the negative eigenvalue. The third condition is important in order to compare different EW.
\begin{definition}
A hermitian operator $W_{\rho_{AB}}$ is an optimal EW for the density operator $\rho_{AB}$ if
\begin{equation}
Tr(W_{\rho_{AB}}\rho_{AB}) \leq Tr(W\rho_{AB})
\end{equation}
for every EW $W$.
\end{definition}
Although the above definition of OEW is different from the one introduced in \cite{7}, the optimal EW of both criteria are equal. According to \cite{7}, W is optimal iff for all $ P \geq 0$, $W'= (1 + \epsilon)W - \epsilon P$ is not an EW. 

\section{ROBUST SEMIDEFINITE PROGRAMS}

In this section we will express the search of an optimal EW for an arbitrary state $\rho_{AB}$ in terms of a robust semidefinite program (RSDP). A semidefinite program (SDP) consists of minimizing a linear objective under a linear matrix inequality (LMI) constraint, precisely, 
\begin{center}
minimize $c^{\cal y}x$ subject to
\end{center}
\begin{equation}
F(x) = F_{0} + \sum_{i=1}^{m}x_{i}F_{i} \geq 0
\end{equation}
where $c \in {\cal C}^{m}$ and the hermitian matrices $F_{i} = F_{i}^{\cal y} \in {\cal C}^{n x n}$ are given and $x \in {\cal C}^{m}$ is the vector of optimization variables. $F(x) \geq 0$ means $F(x)$ is hermitian and positive semidefinite. SDPs are global convex optimization programs and can be solved in polynomial time with interior-point algorithms \cite{8}. For instance, if there are $m$ optimization variables and $F(x)$ is a n\hspace{0.03 cm}x\hspace{0.03 cm}n matrix, the number of operations scales with problem size as $O(m^{2}n^{2})$. SDPs have already been used in different problems of quantum information theory \cite{9} and also in the separability problem \cite{10}. An important generalization of $(3)$ is when the data matrices $F_{i}$ are not constant, i.e., they depend of a parameter which varies within a certain subspace. This family of problems, known as robust semidefinite programs, is given by:
\begin{center}
minimize $c^{\cal y}x$ subject to
\end{center}
\begin{equation}
F(x, \Delta) = F_{0}(\Delta) + \sum_{i=1}^{m}x_{i}F_{i}(\Delta) \geq 0, \hspace{0.2cm} \forall \Delta \in {\cal D}
\end{equation}
where ${\cal D}$ is a given vectorial (sub)space. Note that problem $(4)$ is more difficult to solve than $(3)$, since one must find an optimization vector $x$ such that $F(x, \Delta)$ is positive semidefinite for all $\Delta \in {\cal D}$. One often encounters SDP in which the variables are matrices and in which the inequality depends affinely on those matrices. These problems can be readily put into the form $(3)$ by introducing a base of hermitian matrices to each matrix variable. However, since most of optimization solvers \cite{11} admit declaration of problems in this most general form, it is not necessary to write out the LMI explicitly as $(3)$, but instead make clear which matrices are variables. Equalities constraints involving the optimization variables can also appear in $(3)$ and $(4)$ without any further computational effort. We can now enunciate the main result of this paper.
\begin{theorem}
A state $\rho_{AB} \in {\cal B}(H_{A}\otimes H_{B})$ is entangled, i.e., can not be decomposed as $(1)$ iff the optimal value of the following RSDP is negative:
\begin{center}
minimize $Tr(W\rho_{AB})$ subject to
\end{center}
\begin{equation}
\sum_{i=1}^{d_{A}}\sum_{j=1}^{d_{A}}a_{i}^{*}a_{j}W_{ij} \geq 0, \hspace{0.2 cm} Tr(W) = 1, \hspace{0.2 cm} for\,\,\, all \,\,\, a_{i} \in {\cal C}
\end{equation}
where $d_{A}$ is the dimension of $H_{A}$, $W_{ij} = _{A}\bra{i}W\ket{j}_{A} \in {\cal B}(H_{B})$ and $\ket{j}_{A}$ is an orthonormal base of ${\cal H}_{A}$. If $\rho_{AB}$ is entangled, the matrix $W$ which minimizes $Tr(W\rho_{AB})$ is the OEW for $\rho_{AB}$.
\end{theorem}
{\bfseries proof}: First we have to show that $(5)$ is a genuine RSDP. Note that $W_{ij} = _{A}\bra{i}W\ket{j}_{A}$ and the objective $Tr(W\rho_{AB})$ are both linear in the matrix variable $W$. Thus, $(5)$ can be put into the form $(4)$, where ${\cal D}$ in this case is ${\cal C}^{d_{A}}$. We know that a state $\rho_{AB}$ is entangled iff there exists  an operator $W$ such that $Tr(W\rho_{AB}) \leq 0$ and $_{A}\bra{\psi}\otimes _{B}\bra{\phi}W\ket{\phi}_{B}\otimes \ket{
\psi}_{A}\geq 0$ for all states $\ket{\psi} \in H_{A}$ and $\ket{\phi} \in H_{B}$. Therefore, the matrix $_{A}\bra{\psi}W\ket{
\psi}_{A}$ has to be semidefinite positive for all $\ket{
\psi}_{A} \in H_{A}$. Letting $\ket{
\psi}_{A} = \sum_{j}a_{j}\ket{j}_{A}$, where $\ket{j}_{A}$ is an orthonormal base of ${\cal H}_{A}$, it is straightforward to show that the optimal W given by $(5)$ is the OEW of $\rho_{AB}$. {\bfseries QED}.

In spite of the similarity between $(3)$ and $(4)$, RSDPs  are in general very hard optimization problems. Actually, it was proved that robust semidefinite programs like $(5)$ are NP-HARD \cite{12}.
\begin{corollary}
The determination of the OEW for an arbitrary state $\rho_{AB}$ is a NP-HARD problem.
\end{corollary}

Since $(5)$ is computationally intractable, it is natural to search for approximations of it in terms of SDPs, which are very efficiently solved. These relaxations of RSDP have been intensively studied \cite{13} in the past years and can be classified as probabilistic or deterministic. In this paper we will focus in the deterministic relaxations, where $(4)$ is replaced by an inner convex approximation described by a linear matrix inequality constraint. This inner approximation is then used to find an upper bound to the optimal value of $(4)$. The probabilistic approach, which yields  outstanding results on the separability problem, will be reported elsewhere. As a first example of such relaxations, consider the following adaptation of \cite{14}:
\begin{theorem}
A density operator $\rho_{AB}$ is entangled and the optimal value of $W$ is an EW for it if the result of the following SDP is negative:
\begin{equation}
minimize \hspace{0.2 cm} Tr(W\rho_{AB}) \hspace{0.2 cm}subject \hspace{0.2 cm} to 
\end{equation}
\begin{enumerate}
\item
\begin{center}
$W_{kk} \geq 0, \hspace{0.2 cm} k = 1, 2, .., d_{A}$
\end{center}
\item
\begin{center}
$\frac{1}{d_{A}-1}W_{kk} \pm \frac{\sqrt{2}}{2}(W_{kj} + W_{jk}) \geq 0, \hspace{0.2 cm} 1 \leq k \neq j \leq d_{A}.$
\end{center}
\item
\begin{center}
$\frac{1}{d_{A}-1}W_{kk} \pm \frac{\sqrt{2}}{2i}(W_{kj} - W_{jk}) \geq 0, \hspace{0.2 cm} 1 \leq k \neq j \leq d_{A}.$
\end{center}
\end{enumerate}
\end{theorem}
{\bfseries proof}: Note that:
\[  \sum_{k, j}a_{k}^{*}a_{j}W_{kj} = \sum_{1\leq k\leq j \leq d_{A}}
\left[
 \frac{1}{d_A - 1} |a_{k}|^{2}W_{kk} \right. \]
\[ \left.  + \hspace{0.2cm} a_{k}^{*}a_{j}W_{kj} + a_{j}^{*}a_{k}W_{jk} + \frac{1}{d_{A}-1}|a_{k}|^{2}W_{kk}\right ] \geq 0 \]
Thus, a sufficient condition to $(5)$ is:
\[
M(\ket{e}) = \left[
\begin{array}{cc}
\bra{e}{\frac{W_{kk}}{d_{A}-1}}\ket{e} & \bra{e}{W_{kj}}\ket{e} \\
\bra{e}{W_{jk}}\ket{e} & \bra{e}{\frac{W_{jj}}{d_{A}-1}}\ket{e}
\end{array}
\right ] \geq 0, \hspace{0.2 cm} \forall \ket{e} \in {\cal H}_{B}
\]
This matrix $M(\ket{e})$ is positive semidefinite iff its diagonal entries and determinant are greater than  or equal to zero. From condition 1 it follows that $M_{11} \geq 0$ and $M_{22} \geq 0$. From conditions 2 and 3:
\[ \bra{e}{\frac{W_{kk}}{d_{A}-1}}\ket{e}\bra{e}{\frac{W_{jj}}{d_{A}-1}}\ket{e} \geq 
2[max{\cal f}\bra{e}\frac{\sqrt{2}}{2}(W_{kj} + W_{jk})\ket{e}, \] \[ \bra{e}\frac{\sqrt{2}}{2i}(W_{kj} - W_{jk})\ket{e} {\cal g}]^{2} \geq \bra{e}{W_{kj}}\ket{e}\bra{e}{W_{jk}}\ket{e} \]
$ \Rightarrow det(M(\ket{e}) > 0$. {\bfseries QED}.

We present now the first example of our methodology. We used MATLAB and the package SEDUMI \cite{11} to implement and solve the SDP. 
\subsection{Bell State}
We consider the following Bell state $\ket{\Psi} = (\ket{0}\otimes \ket{0} + \ket{1}\otimes \ket{1})/\sqrt{2}$. It is well known that $\bra{\Psi}W_{\ket{\Psi}\bra{\Psi}} \ket{\Psi} = -1/2$ \cite{bell-otimo}, where $W_{\ket{\Psi}\bra{\Psi}} = -\frac{1}{2}(\ket{00}\bra{11} + \ket{11}\bra{00}) + \frac{1}{2}(\ket{01}\bra{01} + \ket{10}\bra{10})$ is the OEW for this state. Solving the SDP of theorem 2, the following EW was found
\[
W = \left [
\begin{array}{cccc}
0.1057 & 0 & 0 & -0.2887 \\
0 & 0.3943 & 0 & 0 \\
0 & 0 & 0.3943 & 0 \\
-0.2887 & 0 & 0 & 0.1057
\end{array}
\right ]
\]
Since program (6) is only a relaxation of (5), W is not the OEW for $\ket{\Psi}$, $\bra{\Psi}W_{opt} \ket{\Psi} = -0.1835$.

\subsection{Isospectral States}
We now consider the two isospectral matrices:
\[
\rho_{AB} = \left [
\begin{array}{cccc}
1/3 & 0 & 0 & 0 \\
0 & 1/3 & 1/3 & 0 \\
0 & 1/3 & 1/3 & 0 \\
0 & 0 & 0 & 0
\end{array}
\right ] \]
\[
\sigma_{AB} = \left [
\begin{array}{cccc}
1/3 & 0 & 0 & 0 \\
0 & 0 & 0 & 0 \\
0 & 0 & 0 & 0 \\
0 & 0 & 0 & 2/3
\end{array}
\right ]
\]
The positive partial transpose criterion show that $\rho_{AB}$ is entangled, while $\sigma_{AB}$ is separable. Using the SDP $(6)$, we have found the following EW for $\rho_{AB}$:
\begin{equation}
W = \left [
\begin{array}{cccc}
0.1752 & 0 & 0 & 0 \\
0 & 0.1752 & -0.2478 & 0 \\
0 & -0.2478 & 0.0513 & 0 \\
0 & 0 & 0 & 0.5982
\end{array}
\right ]
\end{equation}
where $Tr(W\rho_{AB}) = -0.0313$. The method has also succeeded in the state $\sigma_{AB}$, as the optimal value for $Tr(W\sigma_{AB})$ founded was $2.7330 x 10^{-5}$.

\section{COMPLETE FAMILY OF PARAMETRIZED RELAXATIONS}
\hspace{0.2 cm}It must be stressed that theorem 2 is only one of the possible approximations of (5). In fact, every relaxation of the RSDP constitutes a different method of EW construction and, therefore, a new sufficient criterion of separability. In this section we will show that all these possible relaxations can be parametrized in terms of a family of matrices. However, in order to provide such method, it is necessary firstly to introduce some standard results concerning robust semidefinite programs. One particularly important representation of robust linear matrix inequalities is the Linear Fraction Representation (LFR) \cite{13} \cite{15}. It was showed that every matrix $F \in {\cal C}^{nxc}$ which depends rationally of a varying parameter $\delta \in {\cal C}^{k}$ can be expressed as \cite{15}:
\begin{equation}
F(\delta) = A + B\Delta(I - D\Delta)^{-1}C
\end{equation}
where $A \in {\cal C}^{n x c}$, $B \in {\cal C}^{n x N}, C \in {\cal C}^{N x c}$ and $D \in {\cal C}^{N x N}$ are constant matrices, $r_{1}, ..., r_{k}$ and $N = r_{1}+ ...+ r_{k}$ are integer numbers and $\Delta$ is the following diagonal matrix:
\[ \Delta = Diag(\delta_{1}I_{r_{1}}, ..., \delta_{k}I_{r_{k}}) \]
We can now express problem $(5)$ in terms of a LFR.

\begin{theorem}
A state $\rho_{AB}$ is entangled and the optimal value of $W$ is the OEW for it iff the result of the following RSDP is negative:
\begin{center}
minimize $Tr(W\rho_{AB})$ subject to
\end{center}
\begin{equation}
F(\Delta) = B\Delta(I - D\Delta)^{-1}C > 0, \hspace{0.2 cm} \forall \Delta \in {\cal D}
\end{equation}
where 
\begin{equation}
B = \left [
W_{11} \hspace{0.2cm} ... \hspace{0.2 cm} W_{n1} \hspace{0.2 cm} W_{12} \hspace{0.2cm} ... \hspace{0.2 cm} W_{nn} \hspace{0.2 cm} 0_{d_{A};d_{A}d_{B}}
\right ]
\end{equation}
\begin{equation}
C = \left [
\begin{array}{cc}
0_{d_{A};d_{B}^{2}d_{A}} & L \otimes I_{d_{A}}
\end{array}\right ]^{\cal y}
\end{equation}
\begin{equation}
D = \left[
\begin{array}{cc}
0_{d_{A}d_{B}^{2};d_{A}d_{B}^{2}} & L^{\cal y} \otimes  I_{d_{A}d_{B}}\\
0_{d_{A}d_{B};d_{A}d_{B}^{2}} & 0_{d_{A}d_{B};d_{A}d_{B}}
\end{array}
\right ]
\end{equation}
\begin{equation}
\Delta = Diag\left(
a_{1}I_{d_{A}d_{B}}, \hspace{0.1cm}... , \hspace{0.1cm}a_{d_{A}}I_{d_{A}d_{B}}, \hspace{0.1cm} a_{1}^{*}I_{d_{A}}, \hspace{0.1cm} ... ,  \hspace{0.1cm}a_{d_{A}}^{*}I_{d_{A}}
\right )
\end{equation}
and ${\cal D}$ is the subspace of diagonal matrices in the form of $(12)$, where $a_{j} \in {\cal C}$. $0_{p;q}$ and $I_{p}$ stand for the $p\hspace{0.06 cm}x\hspace{0.06 cm}q$ zero matrix and the $p\hspace{0.06 cm}x\hspace{0.06 cm}p$ identity matrix, respectively. $L$ is an auxiliary matrix given by:
\[ L = [1, 1, ..., 1] \in H_{B} \]
\end{theorem}
{\bfseries proof}: We must show that the LMI of $(8)$ is equivalent to the LMI of $(5)$. In order to do that, we will use constructive formulas of addition and multiplication of LFRs presented in the appendix. Each quadratic term from the LMI of $(5)$ can be written as
\[ a_{i}^{*}a_{j}W_{ij} = {\cal f}0 + 1\times a_{i}^{*}(1 - 0\times a_{i}^{*})^{-1}\times1{\cal g} \times \] 
\[{\cal f}0 + W_{ij}\times a_{i}^{*}(I - 0\times a_{i}^{*})^{-1}\times I{\cal g} \]
A LFR to each term and then to the whole expression can now be obtained using the addition and the multiplication formulas, respectively. {\bfseries QED}.

There are several approximations for robust linear matrix inequalities (RLMI) which are described as LFR \cite{13}. One of particular importance is the Full Block S-Procedure \cite{16}:
\begin{theorem}
(Full Block S-Procedure \cite{16}) The matrix $F(\Delta) = A + B\Delta(I - D\Delta)^{-1}C$ is well posed and satisfies
\begin{equation}
\left [
\begin{array}{c}
I \\ F(\Delta)
\end{array}
\right ]^{\cal y}
\left [
\begin{array}{cc}
0 & X \\ 
X & 0
\end{array}
\right ]
\left [
\begin{array}{c}
I \\ F(\Delta)
\end{array}
\right ] \leq 0 , \hspace{0.3 cm} \forall \Delta \in {\cal D}
\end{equation}
iff there exists a multiplier 
\begin{equation}
P = \left [ \begin{array}{cc}Q & S \\ S^{\cal y} & R \end{array} \right] 
\end{equation}
with
\begin{equation}
\left [
\begin{array}{c}
\Delta \\ I
\end{array}
\right ]^{\cal y}
P\left [
\begin{array}{c}
\Delta \\ I
\end{array}
\right ] \geq 0 , \hspace{0.3 cm} \forall \Delta \in {\cal D}
\end{equation}
such that
\begin{equation}
\left [
\begin{array}{cc}
I & 0 \\ 
A & B \\
0 & I \\
C & D
\end{array}
\right ]^{\cal y}
\left [
\begin{array}{cccc}
0 & X &0 & 0 \\ 
X & 0 & 0 & 0 \\
0 & 0 & Q & S \\
0 & 0 & S^{\cal y} & R
\end{array}
\right ]
\left [
\begin{array}{cc}
I & 0 \\ 
A & B \\
0 & I \\
C & D
\end{array}
\right ] \leq 0
\end{equation}
\end{theorem}
We can now express all possible relaxations of $(5)$ in terms of the multiplier matrix $P$ as follows:
\begin{theorem}
A state $\rho_{AB}$ is entangled iff there exists a multiplier matrix $(15)$ such that $(16)$ and $(17)$ hold, with $X = -I$. The matrices appearing in the LMI $(17)$ are given by equations $(10 - 12)$ and the matrix $\Delta$ is given by $(13)$.
\end{theorem}
{\bfseries proof}: Noticing that $F(\delta) \geq 0$ is equivalent to $(14)$ if 
$X = -I$, the result follows easily from the application of the Full Block S-Procedure (theorem 4) in the RSDP $(9)$. {\bfseries QED.}

The families of matrices P such that $(16)$ is satisfied parametrize all possible relaxations of $(5)$. Although the determination of all such matrices is not a trivial problem, it is a lot easier than $(14)$ and it is current subject of intensive research. Further information on possible choices of the matrix P can be found in \cite{16}. As an example, we consider now the most simple family of matrices P for which $(16)$ holds. In quantum mechanics one usually deals with normalized states $\braket{\psi}{\psi} = 1$. Therefore, the matrix $\Delta$ satisfies $\Delta^{\cal y} \Delta < I$. In this case, the following matrix P gives an approximation of $(5)$:
\[
P = \left [ \begin{array}{cc}
-I & 0 \\
0 & I
\end{array}
\right ]
\]

\section{MULTIPARTITE ENTANGLEMENT}

So far we have only considered the bipartite case. In this section we generalize the previous results  to multipartite states. A density operator $\rho_{1...n} \in {\cal B}(H_{1} \otimes ... \otimes H_{n})$ is separable if it can be decomposed as:
\begin{equation}
\rho_{1...n} = \sum_{i=1}p_{i} \ket{\psi_{i}}_{1} {}_{1}\bra{\psi_{i}}\otimes ... \otimes \ket{\psi_{i}}_{n} {}_{n}\bra{\psi_{i}}
\end{equation}
Since the set of multipartite mixed separable states is also convex, it is possible to apply the Hahn-Banach theorem and establish the concept of EW in a straightforward manner \cite{17}. 
\begin{theorem}
A state $\rho_{1...n} \in {\cal B}(H_{1} \otimes ... \otimes H_{n})$ is entangled, i.e., can not be decomposed as $(18)$ iff the optimal value of the following RSDP is negative:
\begin{center}
minimize $Tr(W\rho_{1...n})$ subject to
\end{center}
\begin{equation}
\sum_{i_{1}=1}^{d_{n}}\sum_{j_{1}=1}^{d_{n}} ... \sum_{i_{n-1}=1}^{d_{n}}\sum_{j_{n-1}=1}^{d_{n}} \left( a_{i_{1}}^{*}...\hspace{0.05 cm} a_{i_{n-1}}^{*} a_{j_{1}}... \hspace{0.05 cm}a_{j_{n-1}} \right.
\end{equation}
\[ \left. W_{i_{1}... \hspace{0.01 cm}i_{n-1}j_{1}... \hspace{0.01 cm}j_{n-1}} \right) \geq 0 \]
\[ Tr(W) = 1, \hspace{0.2 cm} \forall a_{i_{k}} \in {\cal C},\hspace{0.2 cm} 1 \leq k \leq n \]
where $d_{n}$ is the dimension of $H_{n}$, $W_{i_{1}...\hspace{0.01 cm}i_{n-1}j_{1}...\hspace{0.01 cm}j_{n-1}} =  {}_{1}\bra{i} \otimes ... \otimes {}_{n-1}\bra{i}W\ket{j}_{n-1} \otimes ... \otimes \ket{j}_{1} \in {\cal B}(H_{1} \otimes ... \otimes H_{n-1})$ and $\ket{j}_{k}$ is an orthonormal base of ${\cal H}_{k}$. If $\rho_{1...n}$ is entangled, the matrix $W$ which minimizes $Tr(W\rho_{1...n})$ is the OEW for $\rho_{1...n}$.
\end{theorem}
{\bfseries proof}: We know that a state $\rho_{AB}$ is entangled iff there exists an operator $W$ such that $Tr(W\rho_{1...n}) \leq 0$ and $_{1}\bra{\psi}\otimes ... \otimes _{n}\bra{\psi}W\ket{\psi}_{n}\otimes ... \otimes \ket{\psi}_{1}\geq 0$ for all states $\ket{\psi}_{k} \in H_{K}$. Thus, the matrix $_{1}\bra{\psi}\otimes ... \otimes _{n-1}\bra{\psi}W\ket{\psi}_{n-1}\otimes ... \otimes \ket{\psi}_{1}\geq 0$ has to be semidefinite positive for all $\ket{\psi}_{k} \in H_{K}$. Letting $\ket{
\psi}_{k} = \sum_{j}a^{k}_{j}\ket{j}_{k}$, where $\ket{j}_{k}$ is an orthonormal base of ${\cal H}_{k}$, it is straightforward to show that the optimal W given by $(19)$ is the OEW of $\rho_{1...n}$. {\bfseries QED}.

Relaxations for $(19)$ can be obtained using the same arguments exposed before. Since the RLMI of $(19)$ is polynomial in the varying parameters, it can be expressed as a LFR and we can apply the Full Block S-Procedure to the multipartite case. Therefore, all possible deterministic approximations of $(19)$ can also be parametrized by the matrix P. Further results concerning the application of possible families of parameterizations of P in the optimization of EW will be reported elsewhere.

\section{CONCLUSION}
In this paper we have introduced, in the context of Quantum Information, a new class of optimization programs (RSDP), showing that the determination of the OEW for an arbitrary state is NP-HARD. Several possible deterministic approximation scenarios have been proposed to it, yielding  new sufficient criteria of separability. Our results were then straightforwardly generalized to multipartite states. It was also showed that all sufficient criteria of separability might be parametrized by a matrix which satisfies a much simpler linear matrix inequality. Therefore, it is of great importance a systematic study of all possibles families of parameterizations for this matrix.
\section{APPENDIX: COMBINATION OF LFRS}
We provide in this appendix some simple combination rules for addition and multiplication of LFR used in this paper. Consider two matrix described by the LFR format:
\[
F_{i}(\delta) = A_{i} + B_{i}\Delta_{i}(I - D_{i}\Delta_{i})^{-1}C_{i}
\]
The sum of $F_{1}$ and $F_{2}$ has the LFR:
\begin{equation}
F(\delta) = A + B\Delta(I - D\Delta)^{-1}C
\end{equation}
with
\[
A = A_{1} + A_{2}, \hspace{0.2cm} B = [B_{1} \hspace{0.2cm} B_{2}], \hspace{0.2cm} C = [C_{1} \hspace{0.2cm} C_{2}]^{\cal y}
\]
\[
 \hspace{0.2cm} D = Diag(D_{1}, D_{2}), \hspace{0.2cm} \Delta = Diag(\Delta_{1}, \Delta_{2})
\]
The product of $F_{1}$ and $F_{2}$ is given by (20) with
\[
A = A_{1}A_{2}, \hspace{0.2cm} B = [B_{1} \hspace{0.2cm} A_{1}B_{2}], \hspace{0.2cm} C = [C_{1}A_{2} \hspace{0.2 cm} C_{2}]^{\cal y}
\]
\[
 \hspace{0.2cm} D = \left [
\begin{array}{cc}
D_{1} & C_{1}B_{2} \\
0 & D_{2}
\end{array}
\right ], \hspace{0.2cm} \Delta = Diag(\Delta_{1}, \Delta_{2})
\]

 \begin{acknowledgments}

Partial financial support from the Brazilian agencies CNPq,  Institutos do Mil\^enio-Informa\c{c}\~ao Qu\^antica and  FAPEMIG.

\end{acknowledgments}

%\vspace{0.3 cm}

%\hline

%\vspace{0.3 cm}

\end{document}